# New Discoveries in Cosmology and Fundamental Physics through Advances in Laboratory Astrophysics

Submitted by the

American Astronomical Society Working Group on Laboratory Astrophysics
http://www.aas.org/labastro/

Nancy Brickhouse - Harvard-Smithsonian Center for Astrophysics
nbrickhouse@cfa.harvard.edu, 617-495-7438

John Cowan - University of Oklahoma
cowan@nhn.ou.edu, 405-325-3961

Paul Drake - University of Michigan
rpdrake@umich.edu, 734-763-4072

Steven Federman - University of Toledo
steven.federman@utoledo.edu, 419-530-2652

Gary Ferland - University of Kentucky
gary@pa.uky.edu, 859-257-879

Adam Frank - University of Rochester
afrank@pas.rochester.edu, 585-275-1717

Eric Herbst - Ohio State University
herbst@mps.ohio-state.edu, 614-292-6951

Keith Olive[*] - University of Minnesota
olive@physics.umn.edu, 612-624-7375

Farid Salama - NASA/Ames Research Center
Farid.Salama@nasa.gov, 650-604-3384

Daniel Wolf Savin[*] - Columbia University
savin@astro.columbia.edu, 1-212-854-4124,

Lucy Ziurys – University of Arizona
lziurys@as.arizona.edu, 520-621-6525

[*] Co-Editors

# 1. Introduction

As the Cosmology and Fundamental Physics (CFP) panel is fully aware, the next decade will see major advances in our understanding of these areas of research. To quote from their charge, these advances will occur in studies of "the early universe, the microwave background, the reionization and galaxy formation up to virialization of protogalaxies, large scale structure, the intergalactic medium, the determination of cosmological parameters, dark matter, dark energy, tests of gravity, astronomically determined physical constants, and high energy physics using astronomical messengers."

Central to the progress in these areas are the corresponding advances in laboratory astrophysics which are required for fully realizing the CFP scientific opportunities within the decade 2010-2020. Laboratory astrophysics comprises both theoretical and experimental studies of the underlying physics which produce the observed astrophysical processes. The 5 areas of laboratory astrophysics which we have identified as relevant to the CFP panel are atomic, molecular, plasma, nuclear, and particle physics.

Here, Section 2 describes some of the new scientific opportunities and compelling scientific themes which will be enabled by advances in laboratory astrophysics. In Section 3, we provide the scientific context for these opportunities. Section 4 briefly discusses some of the experimental and theoretical advances in laboratory astrophysics required to realize the CFP scientific opportunities of the next decade. As requested in the Call for White Papers, Section 5 presents four central questions and one area with unusual discovery potential. Lastly, we give a short postlude in Section 6.

## 2. New scientific opportunities and compelling scientific themes

Some attempts to unify gravity with the strong and electroweak forces suggests the possibility for temporal and spatial variations of the fundamental "constants" such as the fine-structure constant $\alpha = e^2/\hbar c$ and the electron/proton mass ratio $\mu = m_e/m_p$ (Uzan 2003). Astrophysical research is uniquely situated to address the question of whether these constants are so or if they vary over the lifetime of the universe.

Measuring cosmological parameters has reached amazing levels of precision. Data from the angular power spectrum of the CMB can be used to determine several key cosmological parameters including relative densities of baryonic matter, dark matter, and dark energy as well as test models of inflation (Wong et al. 2008). Observations can also constrain the specific entropy of the universe (Sunyaev & Chluba 2008), the primordial helium abundance (Peimbert et al. 2008), and the CMB monopole temperature (Hamann & Wang. 2008). Studies in these areas lie on the cutting edge of cosmology.

Cosmologists want to probe the structure and reionization of the universe during the dark ages, the period between recombination and the formation of the first stars (Pritchard & Loeb 2008). Such observations are expected to begin in the next decade.

There is now a multitude of evidence for the existence of non-baryonic dark matter. In addition to the above mentioned CMB measurements, there is a host of astronomical observations including the measurement of galactic rotation curves, X-ray emission from clusters of galaxies, and gravitational lensing. All observations are consistent with dark matter comprising roughly 20% of the total energy density in the Universe. Numerous laboratory experiments are currently actively searching for the direct detection of dark matter. A positive detection from these experiments coupled with future data from the Large Hadron Collider (LHC) at CERN may determine the identity of the dark matter in addition to firmly establishing its existence.

Studies of the IGM and the Lyman α forest can be used to constrain the spectral shape and history of the metagalactic radiation field, the chemical evolution of the universe, and the initial mass function of the earliest generation of stars.

### 3. Scientific context

In the standard model of electroweak interactions, fermion masses as well as the weak gauge boson masses are generated through the Higgs mechanism. The standard model contains a fundamental scalar field, the Higgs boson, with self-interactions which allow for the dynamical generation of a non-zero vacuum expectation value. As a consequence, all particles, interacting with the Higgs boson, acquire mass. In an analogous manner, unified theories such as string theory contain additional scalar fields for which there is no fixed background value. If these fields couple to electric and magnetic fields, their background value, which may evolve in an expanding universe, determine the gauge coupling constants, such as the fine structure constant. In these unified theories, variations of the gauge couplings may lead to variations of Yukawa couplings (the couplings of the Higgs bosons to fermions) as well as the variation of the scale at which strong interactions become non-perturbative $\Lambda_{QCD}$. As a result, in addition to variations in $\alpha$, variations in the ratios of particle masses such as $m_e/m_p$ may occur.

Quasar absorption spectral observations have been used to probe for variations in fundamental constants (King et al. 2008, Murphy et al. 2008). Variations in $\alpha$ can be determined from detailed atomic and molecular absorption features. One particularly sensitive test is the many-multiplet method based on the relativistic corrections to atomic transitions using several lines from various elemental species and allows for sensitivities which approach the level of $10^{-6}$. Similarly, measurements for the variability of $\mu$ rely on Lyman and Werner transitions of the $H_2$ molecule.

While the standard model of cosmology can be simply formulated in the context of general relativity, a precise description relies on several key parameters which include the relative densities of baryonic and dark matter, as well, as the dark energy, the rate of expansion of the universe characterized by the Hubble parameter, the shape of the angular power spectrum, the optical depth to the last scattering of the CMB, as well as others. *WMAP* has achieved an enormous leap in the accuracy of these parameters and with the launching of *Planck*, the third generation CMB satellite, a new level of precision will be achieved for CMB temperature and polarization anisotropies (Wong et al. 2008). These measurements can be used to constrain various models for inflation.

CMB observations at decimeter wavelengths are planned for the near future. Any detected CMB distortions will provide an additional means to determine some of the key parameters of the universe such as the specific entropy of the universe, the CMB monopole temperature, and the primordial helium abundance (Sunyaev & Chluba 2008).

Plans have been made to probe the dark ages through observations of the redshifted 21-cm hyperfine line in neutral hydrogen (Bowman et al. 2007). The first generation experiments to carry out these observations are currently under construction.

The first indication for dark matter came from the observations of high velocities of galaxies within clusters of galaxies and required an additional source of gravity beyond that which could be accounted for from light-producing galaxies in the cluster. Similarly, the rotation of stars and gas in spiral galaxies also pointed to the notion that galaxies were embedded in a large and massive halo of dark gravitating matter. The presence of hot X-ray emitting gas also requires a significant amount of gravity to prevent the hot gas from flying off into space. Observations of distant galaxies along the line of sight of a large cluster of galaxies show clear signs of the gravitational lensing of the light. Indeed, recent claims to direct evidence for dark matter came from observations of a collision of clusters of galaxies showing lensing which is directly associated with (non-dissipative) dark matter in tact with the two clusters whereas the (dissipative) hot gas has been stripped from the cluster by the collision (Clowe 2006).

The standard model of strong and electroweak interactions has been very well established by precision measurements at LEP and SLAC. At the energies achieved so far, there are precious few discrepancies which can be taken as a sign of new physics beyond the standard model. Of course a key missing constituent of the standard model is the Higgs boson. Its discovery is the primary goal for the LHC which will begin operations this year. Nevertheless, particle physics models beyond the standard model are known to possess particle candidates for the dark matter. For example, axions are particles which appear in models which attempt to solve the so-called strong CP problem associated with the strong nuclear force.

One problem associated with the presence of fundamental scalar particles in the standard model is known as the hierarchy problem. Put simply, in the context of a unified theory with physics at a very high energy scale, it is very difficult to maintain the scale of the breakdown of the electroweak theory to energies as low as 100 GeV. An elegant solution to this problem is known as supersymmetry which is a theory relating particles which differ in their spin statistics by a ½ integer (in units of $\hbar$). Each known particle in the standard model is associated with a superpartner, yet to be discovered. The discovery of superpartners is also one of the main goals for the LHC. The unification of running gauge couplings at high energy in unified theories also requires an extension of the standard model similar to the supersymmetric extension. The lightest of all the new supersymmetric particles is expected to be stable as its decay is protected by the conservation of a new quantum number in much the same way that the proton is protected by the conservation of baryon number or the electron by the conservation of electric charge. This new massive particle is an ideal candidate for dark matter.

The planned dramatic increase in collecting area for the next generation of ground-based observatories, going from 10 m to 30 m telescopes, is expected to revolutionize observational and theoretical studies of the IGM and the Lyman α forest. Such studies will dramatically improve our ability to constrain the spectral shape and history of the metagalactic radiation field, the chemical evolution of the universe, and the initial mass function of the earliest generation of stars.

## 4. Required advances in Laboratory Astrophysics

Advances particularly in the areas of atomic, molecular, plasma, nuclear, and particle physics will be required for the scientific opportunities described above. Here we briefly discuss some of the relevant research in each of these 5 areas of laboratory astrophysics. Experimental and theoretical advances are required in all these areas to fully realize the CFP scientific opportunities of the next decade.

### 4.1. Atomic Physics

Spectroscopic observations of quasar absorption systems to search for variations in the fundamental constants require accurate transition frequencies and their dependence on these constants for many different atomic systems. Strong systematic uncertainties arise from inaccurate laboratory measurements and some systems of particular importance include ions of Ti, Mn, Na, C, and O (Berengut et al. 2004). Atomic clocks and laboratory transition frequency determinations can also be used to constrain variations in the fundamental constants (Lea 2007).

Interpreting observations from *Planck* will require a precise knowledge of the cosmological recombination history. The complete lack of understanding of the profile of the CMB last-scatter surface dominates the error budget for calculating the cosmological power spectrum. Reducing these uncertainties requires high precision calculations for the cosmological recombination process which in turn rely on accurate data for hydrogen and helium recombination and related processes (Wong et al. 2008).

Observations of the cosmological recombination spectrum can in principle be used to measure the specific entropy of the universe, the primordial He abundance, and the CMB monopole temperature. Interpreting these observations will also require accurate atomic data for recombination spectra in hydrogen and helium and particularly for two photon processes (Sunyaev & Chluba 2008).

Interpreting the planned red-shifted 21-cm observations will require a detailed model of the underlying atomic processes affecting the population of the hyperfine levels. These include resonant scattering of Lyman-alpha photons, spin changing collisions of the hyperfine levels, and related processes (Pritchard & Loeb 2008).

Observations of H II regions are used to infer primordial helium abundances to test the standard model for big bang nucleosynthesis (BBN). Reliable recombination rate coefficients are needed for the observed H I and He I lines as well as for ions of O, Ne, Si, and S (Peimbert et al. 2007).

Interpreting spectra of the IGM Lyman α forest is carried out using both single-phase models and cosmological models of the IGM employing semianalytical approximations for hydrodynamical simulations. These various models used different approximations and assumptions. However, one thing they all have in common is the need to calculate the ionization structure of the photoionized IGM. This is typically carried out using plasma codes written specifically for modeling the ionization structure of photoionized gas. One of the most commonly used codes for this purpose is CLOUDY (Ferland et al. 1998). Fundamental to the accuracy of such plasma codes and any inferred astrophysical conclusions is calculating the correct ionization balance. This requires reliable photoionization, radiative recombination, dielectronic recombination, and charge transfer data (Savin 2000). Accurate data are also needed for photoabsorption (photoexcitation), electron impact excitation, and radiative lifetimes to generate accurate model spectra.

### 4.2. Molecular Physics

Quasar observations of molecular absorption lines can be used to constrain temporal and spatial variations in fundamental constants. For this accurate transition frequencies and their dependence on these constants for many different molecular systems are needed. Some of the more important systems include $H_2$ (King et al. 2008) and OH (Kanekar et al. 2005). Molecular clocks and laboratory transition frequency studies can also be used to constrain variations in the fundamental constants (Lea 2007).

### 4.3. Plasma Physics

Plasma physics at high energy density can produce photoionized plasmas that can benchmark models of the IGM and the reionization of the universe. This is accomplished by producing an intense x-ray source that can irradiate a plasma volume containing relevant species, and measuring the ionization balance and other properties that result.

Work in this direction has begun (Foord et al. 2006; Wang et al. 2008). Much more will be possible in the coming decade as higher-energy facilities come online.

### 4.4. Nuclear Physics

A critical intersection between cosmology and laboratory nuclear astrophysics comes from the role of neutrino mass. Neutrinos are the one identified component of the dark matter, though laboratory bounds and cosmological observations rule out neutrinos as the principal component. Their effect on cosmological evolution, however, is not neglible and oscillation experiments place a lower bound on the contribution. This is important because cosmology can determine the sum of the neutrino masses, a quantity that cannot be extracted from oscillation experiments, which test only $m^2$ differences. The mass scale, in cosmological analyses, is correlated with the parameters of the dark energy equation of state and other neutrino-physics inputs, such as the possibility of additional sterile neutrinos. Thus an independent determination of this scale would be significant.

The best laboratory limit on the sums of the neutrino masses comes from tritium beta decay in combination with oscillations, yielding a bound of 6.6 eV. The inadequacy of this limit is apparent from the more model dependent, but more stringent, result from cosmological analyses of about 0.7 eV. A new tritium endpoint experiment is underway (KATRIN) that is expected to improve this laboratory limit by an order of magnitude, thus providing a check on the current cosmological bound. There are also ambitious new double beta decay experiments that probe, somewhat less directly, the neutrino mass scale. These experiments - carried out by groups in both nuclear and particle physics - have much greater reach and could potentially establishing bounds on the order of 0.05 eV, if mounted at the requested one-to-ten-ton scale. It is anticipated that the cosmological bound will also improve rapidly: analyses based on PLANCK plus SDSS and on CMBpol plus SDSS may be able to reach 0.21 and 0.13 eV, respectively. With future data from weak lensing surveys, this bound could move below 0.1 eV.

The concordance between laboratory and cosmological neutrino mass determinations could be of great importance. One can envision a cosmological mass determination and measurement of the currently unknown third mixing angle (the subject of current reactor and long-baseline programs) in the next decade. In total there would be six constraints (two mass differences, three mixing angles, the mass scale) imposed on the double beta decay Majorana mass. A lack of concordance would yield new physics, e.g., constraining the relative CP of the mass eigenstates and/or their Majorana phases.

The detection of neutrinoless double beta decay would have another important implication for cosmology: it is the only practical low-energy tool for establishing lepton number violation and Majorana masses, ingredients in theories that invoke leptogenesis as a mechanism for generating the cosmological baryon number asymmetry.

The first stars are thought to have formed ~ 100 Myr after the Big Bang: dark matter halos of $10^{5-6}$ solar masses at redshift $z \sim 30$ attract enough primordial gas to produce significant baryon densities and temperatures. By $z \sim 20$ sufficient cooling has occurred to produce gravitational instability. The first stars played an important role in the evolution of the early universe. Prior to their formation the universe consisted mostly of neutral hydrogen and helium, and was dark due to the absorption of light by these atoms. The ultraviolet photons from these early stars helped reionize the early universe, ending the "cosmic dark age". These stars thus influenced conditions for early galaxy formation (e.g., the first 1 Gyr), through their radiation, heavy element production, kinetic energy input, and generation of shock waves and electromagnetic fields.  This epoch is interesting cosmologically because it is accessible: WMAP and studies of distance quasars by SDSS have provided some information on the universe's reionization history, while VLT, Subaru, and Keck have helped us understand the early evolution of metals.

Nuclear physicists are involved in this interdisciplinary field through their efforts to understand the structure, very rapid evolution, and death of these massive, extremely low metalicity stars - and how this evolution affects the interstellar medium.  The work includes the evolution of progenitor models and large-scale numerical simulations to understand the mechanisms for their deaths (e.g., pair instability supernovae vs. collapse to a black hole).  They are also very interested in understanding the patterns of the metals that might emerge from the ejecta - a problem that connects observations of metal-poor stars to laboratory measurements of nuclear masses and decays.

Nuclear transitions are possible candidates for measuring the time variation in α.  One system of particular interest is $^{229}$Th (Litvinova et al. 2009).  Such studies will require highly accurate nuclear model calculations.

**4.5 Particle Physics**

The search for dark matter has become a multi-faceted research enterprise. The search for candidates associated with supersymmetric theories exists on three fronts: accelerator searches, direct detection searches and indirect detection searchs.  Accelerator searches are geared towards new particle discoveries. Supersymmetry predicts many new massive particles and the discovery of any of these particles would go a long way in our understanding of the possible nature and identity of dark matter.  However, as it is expected that the dark matter is neutral under strong and electromagnetic interactions, it is not likely that the LHC will detect the dark matter; rather this new particle would be expected to escape the detectors and appear as missing energy much like neutrinos.  Similarly, indirect detection experiments look at by products of interactions of dark matter.  For example, dark matter annihilation in the galactic halo could provide unique signatures found in the gamma ray, positron, or antiproton backgrounds.  Many experiments are currently running looking for these signatures.  Most notable perhaps is Fermi, mapping out the gamma ray sky.  Another potential signature comes from the scattering and eventual trapping of dark matter particles in the sun.  Annihilations in the sun, would produce high energy neutrinos from the sun and provide a unique signature for dark matter.  The Amanda and IceCube detectors active at the South Pole are geared for the indirect detection of dark matter.

In contrast to accelerator and indirect detection searches, direct detection experiments are key to the actual discovery of dark matter.  These laboratory experiments look for the elastic scattering of dark matter particles with nuclei in low background environments located deep underground. There has been dramatic improvement in the sensitivity of these detectors over the last several years. These detectors are designed to look for weakly interacting massive particles (WIMPS) with masses between 10 GeV and 10 TeV through the nuclear recoil (with energies between 1 and 100 keV) of an elastic scattering event. There are several running experiments with detectors consisting of a wide range of nuclear targets. The best limits on the elastic scattering cross section come from two experiments, CDMS and XENON10.  CDMS, currently located in the Soudan mine in Minnesota, uses cryogenic Ge and Si detectors. XENON10 , located at  Grand Sasso, uses liquid Xe as its active component.  Current limits are at the level of a few times $10^{-8}$ pb. Both experiments have planned upgrades.

Terrestrial experiments search for axions by looking for their conversion to photons in a strong magnetic field (Sikivie 1983). Two high "Q" cavities are currently running. The ADMX experiment at LLNL uses electronic amplifiers with low noise temperature to enhance the conversion signal. Currently this experiment has set limits on the axion mass, excluding masses between 1.9 and 3.3 μeV if the axion constitutes the dark matter. This experiment is being upgraded with the use of SQUID amplifiers. A second experiment CARRACK in Kyoto uses highly excited atoms to detect microwave photons resulting from the axion conversion. This experiment excludes masses around 10 μeV and is being upgraded to a sensitivity capable of probing masses between 2 and 50 μeV.

## 5. Four central questions and one area with unusual discovery potential

### 5.1 Four central questions

- Are the fundamental "constants" truly constant?
- How accurately can we measure the cosmological parameters?
- What is the nature and identity of the dark matter?
- What is the shape and history of the metagalatic radiation field, the chemical evolution of the universe, and the initial mass function for the first stars?

### 5.2 One area with unusual discovery potential

- What is the history and structure of the universe during the dark ages?

## 6. Postlude

Laboratory astrophysics and complementary theoretical calculations are the part of the foundation for our understanding of CFP and will remain so for many generations to come. From the level of scientific conception to that of the scientific return, it is our understanding of the underlying processes that allows us to address fundamental questions in these areas. In this regard, laboratory astrophysics is much like detector and instrument development; these efforts are necessary to maximize the scientific return from astronomical observations.